\documentclass[preprint,3p,times]{elsarticle}

\usepackage{amssymb}
\usepackage{bm}
\usepackage{amsmath}
\usepackage{subcaption}
\usepackage{setspace}
\usepackage{color,soul}

\journal{arxiv}
\begin{document}
\begin{frontmatter}

\title{Applicability of Ideal Gas Law in the Air-bearing of Hard Drives}

\author[inst1, inst2]{Roshan Mathew Tom}
\author[inst3]{Sukumar Rajauria}
\author[inst3]{Qing Dai}

\affiliation[inst1]{organization={Department of Mechanical Engineerng, UC Berkeley},
            city={Berkeley},
            postcode={94720}, 
            state={CA},
            country={USA}}
\affiliation[inst3]{organization={The CTO Office, Western Digital Technologies},
            city={San Jose},
            postcode={95119}, 
            state={CA},
            country={USA}}

\affiliation[inst2]{Corresponding Author: roshantom@berkeley.edu}

\begin{abstract}
    This report evaluates the applicability of the ideal gas law in the pressurized air-bearing of hard disk drives when calculating the relative humidity. We employ a semi-analytical numerical method that solves vapor-liquid equilibrium using the Redlich-Kwong and Peng-Robinson equation of state to calculate the saturation pressure of water. The deviation from the ideal gas law is quantified and examined through saturation pressure isotherms. We find that at low temperatures, lighter gases such as helium show little deviation from the ideal gas law, whereas heavier gases such as nitrogen deviate significantly. As temperature increases, the difference between the gases decreases. The results suggest that in areas of low temperature, the non-ideal nature of gases must be taken into account.
\end{abstract}

\begin{keyword}
HAMR \sep Head-Disk Interface \sep Humidity \sep Vapor-Liquid Equilibrium
\end{keyword}

\end{frontmatter}


\section{Introduction}
For many decades, hard disk drives (HDDs) have been a dominant form of data storage, and it continues to do so. Since conventional HDDs are approaching the theoretical limits of areal density, known as the superparamagnetic limit, HAMR (Heat-assisted magnetic recording) promises to be the key to breaking this barrier \cite{Kryder2008}. In HAMR, a tiny laser is used to heat a nanoscale spot on the disk to its curie temperature. This enables packing the bits in smaller grains, increasing the areal density. However, the presence of the laser brings additional complexities to the head-disk interface, and its reliability remains a significant challenge \cite{Kiely2017}. Therefore, studying the characteristics of the air-bearing is essential to building a robust HAMR drive.

One significant aspect of the air-bearing is the humidity. Much research has been published regarding the effect of humidity on the head-disk interface \cite{Karis2005}. It has been shown to induce a water monolayer on the disk surface \cite{Cheng2019, Shukla_2002}, decrease the tribocharge in the slider-disk interface \cite{Lee_2007}, influence the lubricant transfer between the head and the disk \cite{Kim2009}, enhance the heat transfer between the protrusions from the head in contact with the disk when the RH is greater than 75\% \cite{Cheng2020_hum}, affect the spreading rate and chemisorption of the lubricants on the disk surface \cite{Karis_2000}, reduce the vaporization energy and thus increase evaporation of lubricants during thermal desorption \cite{Lei_2000, Tyndall_2001}, and promote corrosion of the recording layers \cite{Dubin_1982}. These results show a clear relationship between humidity and the reliability of Hard disk drives. An underlying assumption is that the ideal gas law applies in the head-disk interface. However, since the pressures can exceed $50$ atmospheres, real gas effects can come into play. The effect of the air-bearing composition is also taken into account. Since the HAMR head-disk interface contains advanced lubricants, tighter spacing controls, and elevated temperatures, it is essential to reexamine these assumptions to build reliable devices. A key parameter that determines the water content is the saturation vapor pressure. Therefore, we evaluate these assumptions by calculating saturation vapor pressure using a non-ideal gas law and comparing them to an ideal gas.

The outline of the paper is as follows. In Section \ref{CalcPressure}, we will introduce the concepts that are used to calculate the change in saturation pressure in the midst of different gases. In Section \ref{VaporPressureRes}, we will analyze the saturation pressure of water under various environments and explain the behavior. Finally, in Section \ref{Concl}, we will conclude and describe our future work.

\section{Assumptions and Methods}
\label{CalcPressure}
Introduced by Gilbert Lewis in 1901 \cite{Lewis1901}, fugacity ($f$) refers to the tendency of a substance to pass from one chemical phase to another. It is related to the chemical potential ($\mu$, molar Gibbs free energy) by the following relation: 
\begin{align}
    \mu = \mu_0 + RT \ln \left(\frac{f}{f_0}\right)   
\end{align}\
where $\mu_0$ and $f_0$ are the chemical potential and fugacity at a reference state. $R$ and $T$ are the gas constant and temperature of the system. Fugacity has units of pressure and is equal to the hypothetical pressure that a real gas needs to have to satisfy the ideal gas law. In a system with multiple substances in liquid/gas phases, the condition of equilibrium are that the fugacity of species $i$ in its gas ($f_i^{gas}$) and the liquid phase ($f_i^{liquid}$) are equal.
\begin{align}
    \label{EquilibriumEq}f_i^{liquid} &= f_i^{gas}
\end{align}
\subsection{Fugacity in gas phase}
The fugacity of a species in the gas phase is given by, 
\begin{align}
    f_i^{gas} &= \phi_i y_i P
\end{align}
where $\phi_i$, $y_i$, and $P$ are the fugacity coefficient, mole fraction of the given species, and the total pressure, respectively. Equation \ref{FugCoeff} can then calculate the fugacity coefficient \cite{Chueh1967}
\begin{align}
    \label{FugCoeff} RT \ln{\phi_i} &= \int_V^\infty \left[ \left(\frac{\partial P}{\partial n_i}\right)_{T,V,n_j} - \frac{RT}{\sum n_iv} \right]dV - RT \ln{z}\\
    & \text{where, }z = \frac{Pv}{ RT}
\end{align}
where, $n_i$, $V$, and $z$ are the mole count of species $i$, molar volume, and the impressibility factor. Solving Eq \ref{FugCoeff} requires a choice of an appropriate EOS. We have chosen the Redlich-Kwong (RK) EOS due to its relative simplicity and accuracy. For a pure gas, the RK equation is
\begin{align}
    \label{RKEq} P &= \frac{RT}{v-b} -  \frac{a}{\sqrt{T}v(v+b)}\\
    a &= \frac{\Omega_a R^2 T_c^{2.5}}{P_c} \\
    b &= \frac{\Omega_b R T_c}{P_c}
\end{align}
where $a$ and $b$ are parameters related to the intermolecular forces and the finite volume of the gas particles, respectively. $T_c$ and $T_p$ are the critical temperature and pressure. $\Omega_{a/b}$ is a species-dependent constant but roughly equals 0.4227 and 0.0867, respectively. In case of mixtures, the parameters $a$ and $b$ may be replaced by effective parameters using appropriate mixing rules. Chueh and Prausnitz prescribed a mixing rule that is applicable to high-pressure cases given by: 
\begin{align}
    a &= \sum_{i = 1}^N \sum_{j = 1}^N y_iy_ja_{ij} \\  
    b &= \sum_{i = 1}^N y_ib_i   
\end{align}
where the subscript $i$ refers to species $i$. The term $a_{ij}$ is given by: 
\begin{align}
    a_{ij} &= \frac{\left( \Omega_{ai} + \Omega_{aj}\right)R^2T_{cij}^{2.5}}{2P_{cij}} \\
    P_{cij} &= \frac{z_{cij}RT_{cij}}{v_{cij}} \\
    v_{cij} &= \frac18 \left( v_{ci}^{1/3} + v_{cj}^{1/3} \right)^3 \\
    z_{cij} &= 0.291 - 0.08 \left( \frac{\omega_i + \omega_j}{2} \right) \\
    T_{cij} &= \sqrt{T_{ci}T_{cj}}(1-k_{ij})
\end{align}
where, $\omega$ is the acentric factor and $k_{ij}$ is a deviation from the geometric mean for $T_{cij}$. It is characteristic of the $i-j$ interaction. Now, if we insert Eq \ref{RKEq} in Eq \ref{FugCoeff}, we get
\begin{align}
    \label{FugFinal}\ln{\phi_i} = \ln{\left( \frac{v}{v-b}\right)} - 2\frac{\sum_j y_ia_{ij}}{RT^{3/2}b}\ln{\left( \frac{v+b}{b}\right)} + \frac{ab_i}{RT^{3/2}b^2} \left[ \ln \left(\frac{v+b}{v}\right) - \frac{b}{v+b} \right] 
    + \left( \frac{b_i}{v-b}\right) - \ln{\left(\frac{Pv}{RT}\right)}
\end{align}
where $v$ is the molar volume of the gas mixture, obtained by solving Eq \ref{RKEq} and taking the largest real root for $v$.

\subsection{Fugacity of the liquid phase}
The fugacity of a liquid phase is given by \cite{Chueh1968}: 
\begin{align}
    f^{liquid}_i &= \gamma_i x_i f_i^o \exp{\left( \frac{\Delta P v_i^l}{RT} \right)}
\end{align}
where $\gamma$ is the activity coefficient, $x_i$ is the mole fraction in the liquid phase, and $f_i^0$ is the reference fugacity at a known pressure. The exponential term is the change in fugacity due to the ambient pressure and is known as the Poynting correction. $\Delta P$ is the change in pressure and $v_i^l$ is the molar volume of the species in its liquid state. Since we are operating in temperatures well above the critical temperatures for the gases in the air-bearing (Nitrogen, Helium, Oxygen), we will consider these gases as non-condensable. Further, since the dissolved gases do not significantly change the mole fraction of liquid water, we will approximate the liquid to be made of pure water, which implies $x_w = 1$ and $\gamma_w = 1$. Therefore, for liquid water
\begin{align}
    \label{FugWater}f^{liquid}_w &= f_w^o \exp{\left( \frac{\Delta P v_w^l}{RT} \right)}
\end{align}
where the subscript $w$ refers to water. Now, the reference fugacity of liquid water can be readily calculated using Eq \ref{FugFinal} and invoking Eq \ref{EquilibriumEq} in a vacuum using the known experimental value of the saturation vapor pressure. The molar volume can also be experimentally taken or derived from an appropriate equation of state. In our case, we used the Peng-Robinson EOS, which is said to predict the liquid density accurately \cite{Mathias1989}. 

\subsection{Calculating Saturation Pressure}
\label{csp}
The procedure to calculate the saturation pressure of water at pressure $P$, starts with an estimate of the vapor pressure at ambient pressure. We use the Lee-Kesler \cite{LeeKesler1975} method which is given by, 
\begin{align}
    \ln{P_{r}}  &= f_0 + \omega f_1 \\
                \;\;\;\;\;\;\;\;\;\;\; \;\;\;\;\;\;\;\;\;\;\;\;\;\;& f_0 = 5.92714 - \frac{6.09648}{T_r} - 1.28862\log(T_r) + 0.169347T_r^6;\\
                & f_1 = 15.2518 - \frac{15.6875}{Tr} - 13.4721\log(T_r) + 0.43577T_r^6; \\
                & P_r = \frac{P_v}{P_c}, T_r = \frac{T}{T_c}
\end{align}
where $T_r$, $T$, and $T_c$ refer to the reduced temperature, ambient temperature, and critical temperature, respectively. $P_v$ and $P_c$ refer to the saturation pressure and critical pressure, respectively. This method is reliable for pressure around 1 atmosphere where the ideal gas law holds. Then, assuming the gas is purely water vapor, we use Eq \ref{FugFinal} to calculate the reference state fugacity of pure water vapor in equilibrium with liquid water. Then, invoking Eq \ref{EquilibriumEq}, we have the standard state fugacity of liquid water ($f_w^0$). Then, using Eq \ref{FugWater}, we can calculate the fugacity of liquid water at an elevated pressure. 

We have the choice of two boundary conditions, which, in reality, leads to the same results. The elevated pressure can be either due to the air-bearing gas alone or the sum of both the air-bearing gas and water vapor pressure. We assume the latter in our calculation since it is easier to implement. We begin by defining a mixture containing secondary gases at pressure $P$. The RK equation calculates the required moles of each secondary gas. Then, we remove some of these secondary gases in increments, replacing them with water molecules such that the total pressure is maintained at $P$. In each iteration, we calculate the fugacity of the water vapor using Eq \ref{FugFinal}. This process is repeated until we have the fugacity of the water vapor equal to the fugacity of liquid water at elevated pressure calculated earlier (Eq \ref{EquilibriumEq}). The partial pressure of water in this new equilibrium is the saturation vapor pressure of water at pressure $P$. 

\section{Results and Discussion}
\label{VaporPressureRes}
Although hard disks are filled with air or Helium, we will consider gases - Nitrogen, Helium, Argon, and Oxygen- to understand the characteristics of saturation pressure. 10 \% by volume of Oxygen is added to pure Nitrogen, Helium, and Argon to reflect experimental conditions. We will also compare them with a hypothetical ideal gas. Further, in three, Even though the air is mainly a mixture of nitrogen and oxygen, we will continue to use the term air-bearing, even if it is composed of other gases, to maintain convention in this field. The critical values and the RK equation parameters for all the gases are shown in Table \ref{Critical_RKParams}. Then, using the method described in section \ref{csp}, we can calculate the saturation pressure at any temperature and pressure.

\begin{figure}[ht]
    \centering
    \includegraphics[width = \textwidth]{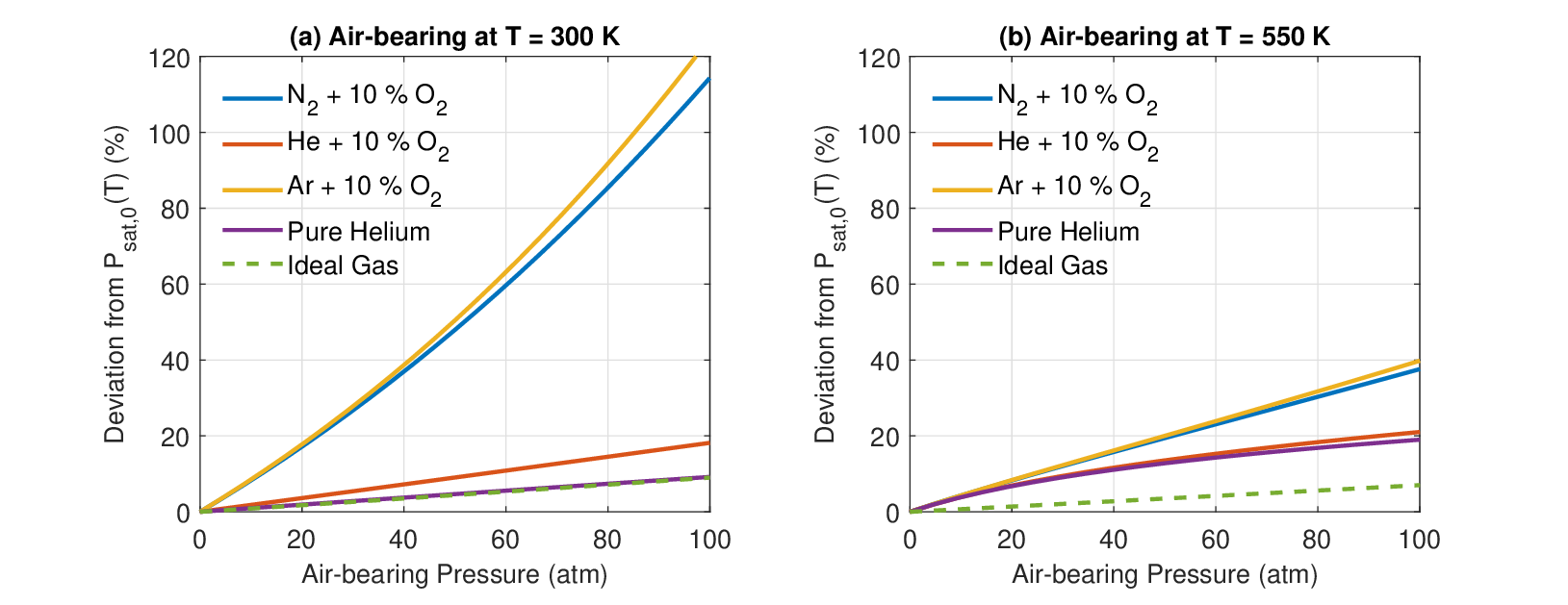}
    \caption{Deviation of saturation pressure of water by increasing the air-bearing pressure. p$_{\text{sat},0}$ is the saturation pressure at 1 atm. (a) Air-bearing temperature is 300 K and (b) Air-bearing temperature is 550 K.}
    \label{SatPressure}
\end{figure}

We plot the deviation of the saturation pressure with the air-bearing pressure (Fig. \ref{SatPressure}). The reference saturation pressure, p$_{\text{sat},0}$ is the saturation pressure at 1 atmospheres and a given temperature. p$_{\text{sat},0}$ for fig. \ref{SatPressure}a and fig. \ref{SatPressure}b are 300 K and 550 K, respectively. At lower temperatures (Fig \ref{SatPressure}a), the immediate observation is that in all mixtures, an increase in pressure results in an increase in the saturation pressure of water. Even in the case of an ideal gas, we observe a deviation of 10\% at 100 atm pressure. This is because the fugacity of liquid water increases with pressure, as Eq \ref{FugWater} would predict. Above this baseline, we observe an increase of more than $100$\% over 100 atm in the case of Nitrogen and Argon. This is because of the intermolecular forces acting with water. The relative magnitude of increase closely follows the parameter $a_{12}$ in Table \ref{Critical_RKParams}, which suggests that the air-bearing gases exert an attractive force on the water molecules, enabling a higher amount of water to exist at a given temperature. Other researchers have also experimentally observed these differences \cite{Mansfield1965, Sechrist1963, Kingdon1963}. At higher temperatures (Fig \ref{SatPressure}b), the deviation is less than $40$\% at 100 atm for all mixtures. This probably is because the molecules are traveling so fast that the effect of the intermolecular forces is masked. Further, we observe a difference in the case of Helium. At room temperature, pure Helium behaves like an ideal gas; however, at higher temperatures, it has almost twice the deviation as the ideal gas. Since the $b_1$ is the only parameter differentiating Helium from an ideal gas, it must also be responsible for this observed deviation. $b_1$ represents the finite volume, and so, Fig. \ref{SatPressure}b suggests that the finite volume plays a significant role at high temperatures. At 300 K, When the pure Helium is mixed with 10\% Oxygen, the net deviation increases from 10\% to 20\% at 100 atm pressure. The intermolecular forces from Oxygen are offsetting the pure Helium's behavior.

Some of the differences are better observed when we plot the deviation of saturation pressure of water over temperature by keeping the pressure constant as in fig \ref{SatPressure}. First, we note that the deviation of ideal gas reduces from 10\% to 5\% over by increasing the temperature. This is mainly because at a fixed $P$, the exponential term in Eq. \ref{FugWater} decreases when we increase the temperature. In the case of heavier gases (Nitrogen and Argon), the deviation rapidly decreases to roughly 10\%. In contrast, in the case of lighter gases (Helium), the deviation somewhat oscillates to roughly 10\%. This is due to the interplay between terms $a_1$ (which corresponds to the intermolecular force), $b_1$ (corresponding to the size), and the temperature. At lower temperatures, intermolecular forces dominate, as seen by the magnitude of deviation of p$_{\text{sat},0}$, whereas at higher temperatures, the finite volume starts to play a role.

\begin{table}[ht]
\centering
\begin{tabular}{ccccccc}
Gas & $T_c$ (K) & $P_c$ (bar) & Acentric Factor $(\omega)$ & $a_{11}$ (L$^2$bar/mol$^2$) & $b_1$ (L/mol) & $a_{12}$ (with Water) \\ \hline
Helium & 10.47 & 6.758 & 0 & 0.0155 & 0.01117 & 0.4953 \\
Hydrogen & 33 & 13 & -0.220 & 0.1423 & 0.01830 & 1.4530 \\
Water & 647 & 220.5 & 0.344 & 14.28 & 0.02115 & 5.1653 \\
Nitrogen & 126 & 34 & 0.04 & 1.550 & 0.02671 & 4.9062 \\
Oxygen & 126 & 34 & 0.04 & 1.550 & 0.02671 & 4.9062 \\
Ideal gas & N/A & N/A & 0 & 0 & 0 & 0 \\
Argon & 151 & 48.7 & 0 & 1.7013 & 0.02235 & 5.1795 \\ \hline
\end{tabular}
\caption{The critical values and RK equation parameters for various gases}
\label{Critical_RKParams}
\end{table}

\begin{figure}[ht]
    \centering
    \includegraphics[width = \textwidth]{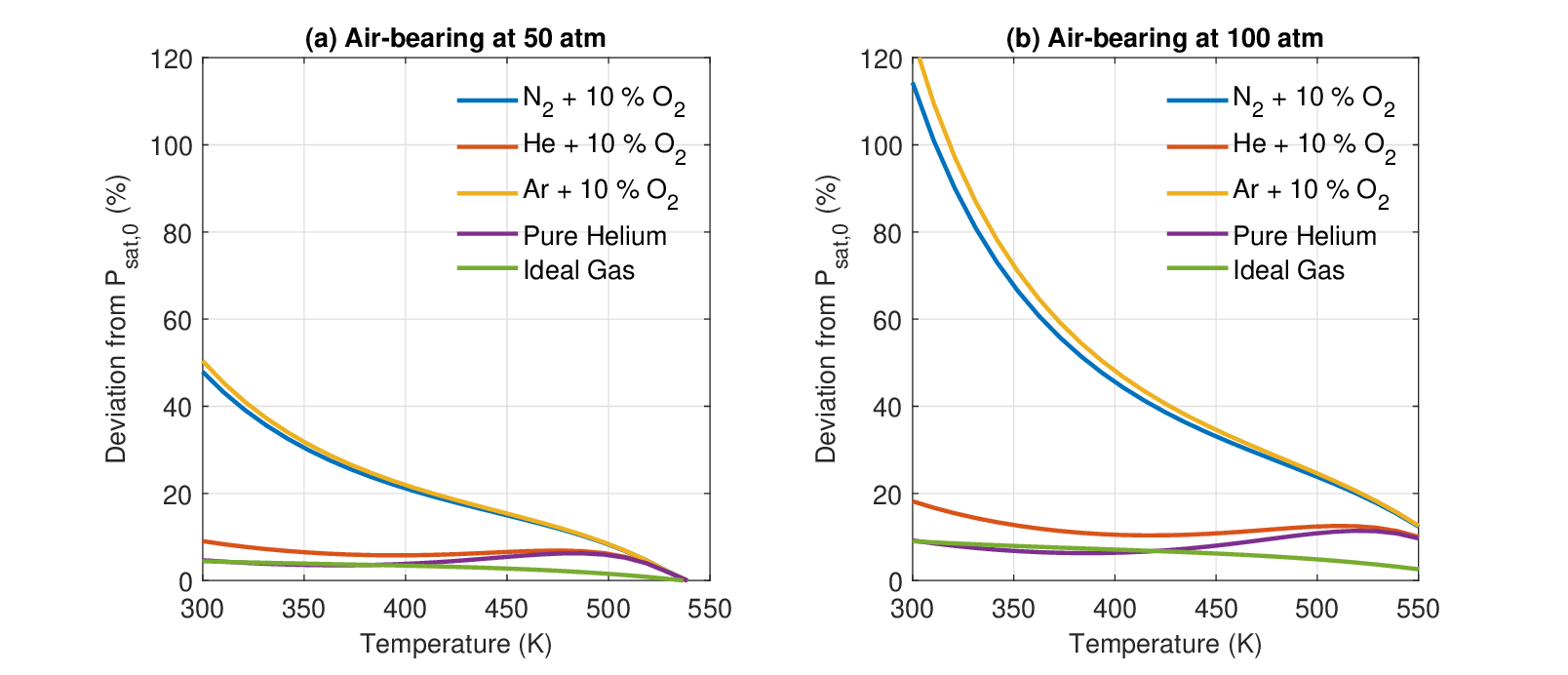}
    \caption{Deviation of saturation pressure of water by increasing the air-bearing pressure to 100 atm. p$_{\text{sat},0}$ is the saturation pressure of water at 1 atm and the given temperature on the x-axis.}
    \label{SatPressure2}
\end{figure}

\section{Conclusion}
\label{Concl}

\par This report presented a procedure to calculate the saturation pressure of water at high-pressure air-bearing using real gas EOS. We observed that at lower temperatures, the saturation pressure significantly deviates from the ideal gas law due to intermolecular force. On the other hand, at higher temperatures, the finite volume of a gas starts to influence the saturation pressure. Nevertheless, the overall significance of the real gas effects reduces at higher temperatures. 
\par In HAMR, the temperatures at the head-disk interface exceed 600 K. If the trend observed in Fig. \ref{SatPressure2} continues, the deviation is predicted to be less than 10\%. Therefore, it may not be relevant to consider real gas effects for HAMR. However, the differences between helium and air may be observed in specific scenarios, especially in cooler but high-pressure air-bearing. A direct experimental procedure to calculate the saturation pressure is necessary to ensure confirmation of the model. However, the results obtained by Manfield \cite{Mansfield1965} and \cite{Sechrist1963} support the model presented here.

\section*{Acknowledgements}
The authors thank Daniel Matsuka and Oscar Ruiz of Western Digital for their helpful discussions.

\bibliographystyle{elsarticle-num} 
\bibliography{cas-refs}

\end{document}